\documentclass{ws-procs975x65}

\begin{document}

\title{FUNCTIONAL APPROACH TO (2+1) DIMENSIONAL
GRAVITY\footnote{\uppercase{P}resented by \uppercase{P. M}enotti}}

\author{L. CANTINI}
\address{Scuola Normale Superiore and INFN, Sezione di Pisa, Italy \\
E-mail: cantini@df.unipi.it}

\author{P. MENOTTI \footnote{\uppercase{W}ork partially
supported by \uppercase{MIUR}.}}
\address{Dipartimento di Fisica della Universit\`a di Pisa and\\ 
INFN, Sezione di Pisa, Italy\\
E-mail: menotti@df.unipi.it}  


\maketitle

\abstracts{
We work out the phase-space functional integral of the gravitational
field in $2+1$ dimensions interacting with ${\mathcal N}$ point
particles in an open universe. 
}
Gravity in $2+1$ dimensions \cite{DJH} has attracted notable interest
both at the classical and at the quantum level. The functional integral
approach to quantum gravity in $2+1$ 
dimensions in absence of matter on closed spaces has been given by 
Carlip\cite{carlip}.
Here  we deal with the quantum treatment of the gravitational field
interacting with ${\mathcal N}$ particles in $2+1$
dimensions on open spaces \cite{cantinimenotti}. Even at the classical
level, in presence of particles the problem acquires a highly non
trivial
dynamics also on open spaces; in the case of open spaces when the
topology is that of the plane the maximally slicing gauge can be adopted,
leading to notable simplifications \cite{BCVW,MS,CMS1}. 

We start with the phase space functional integral known as Faddeev
formula\cite{henneaux} 
\begin{equation}
Z= \int 
\prod_{n=1}^{\mathcal{N}}D[P_n]
D[\pi^{ij}] D[h_{ij}] D[N^i] D[N]
\delta(\chi)\prod_{i=1}^2\delta(\chi^i) 
|{\rm Det}\{\chi^\mu, H_\nu\}| e^{i S}
\end{equation}
where $\chi, \chi^i$ are three gauge fixings.
The space metric $h_{ij}$ is defined on the punctured plane
$R^2\setminus\{q_1,\dots q_{\mathcal N}\}$, where $q_1,\dots
q_{\mathcal N}$ are 
the particle position. Thus integration on the metric
$h_{ij}$ implicitly contains the integration on the particle
positions. 
The integration over $D[N]$ and $D[N^i]$ gives rise to the delta
functions $\delta(\frac{H}{\sqrt{h}})$, $\delta(\frac{H_i}{\sqrt{h}})$ 
in the above equation as discussed in \cite{carlip}. 

The integration measures of the metric and of the conjugate momenta
$\pi^{ij}$ are assumed as those induced by the space DeWitt
distance. The integration 
on the metric is performed by parameterizing the space metric $h_{ij}$
in terms of the particle positions, the conformal factor and
a finite diffeomorphism according to
$h_{ij} = F^*( e^{2\sigma} \delta_{ij})$
being $F$ a 2-dimensional diffeomorphism.
Similarly the conjugate momenta (tangent space to the space metrics)
can be parametrized as 
$
\pi^{ij} = \frac{\pi}{2} h^{ij}+ \pi^{TTij}+ {\sqrt h}(PY^0)^{ij}, 
$
being $Y^0$ the square integrable vector
fields vanishing at the punctures and $\pi^{TTij}$ belongs to the
orthogonal (traceless) 
complement to $(PY^0)^{ij}$. The previously defined $\pi^{TTij}$
can be written as linear combinations of the square integrable
meromorphic  
quadratic differentials
\begin{equation}\label{Qbase1}
Q_{kzz}=\frac{1}{z-z^c_k}-\frac{1}{ z-z^c_1},
~~~~Q_{k\bar z \bar z}=0,~~~~(k=2,\dots {\mathcal N})
\end{equation}
being $z^c_i$ the position of the particles in the conformal gauge.
In the performed changes of integration variables, functional
determinants related to the 
puncture formulation of string theory \cite{dhoker} occur; it is
remarkable that they all cancel out in the final reduction.

While the choice of the maximally slicing gauge (Dirac gauge) $\chi=
\pi=0$
plays a very important role in the reduction of the functional
integral, no trace is left of the space gauge fixings $\chi_1,\chi_2$;  
as a matter of fact one can replace them
by the so called geometric approach \cite{alvarez}, which allows to
factorize the infinite volume of the gauge diffeomorphisms.

A fundamental role in the treatment is played by the boundary term
which is computed through the procedure put forward
in \cite{hawkinghunter} and  
which at the end builds up the reduced hamiltonian. The boundary term
is computed by exploiting the asymptotic behavior of the conformal
factor which classifies the conical nature of the space at infinity
and the solution of the hamiltonian constraint which gives rise the an
inhomogeneous Liouville equation. By exploiting an inequality due to
Picard such boundary term can be computed rigorously
\cite{cantinimenotti} and turns out to
be given by the logarithm of the constant part in the asymptotic
behavior of the conformal factor.

We reach in this way the functional integral
\begin{equation}\label{reduced}
Z=\int \prod_{n=2}^{\mathcal N}D[{z'}^c_n]D[\bar{ z'}^c_n]D[t_n]D[\bar
t_n]~e^{i\int (\sum_{n=2}^{\mathcal N}( t_n\dot {z'}^c_n +
\bar t_n \,\dot\bar{z'}^c_n- H_B)dt}, 
\end{equation}
i.e. all functional determinants cancel out and we reach the same
expression which would have been derived from the quantization of the
reduced classical particle dynamics. The main point in
achieving such a result 
is the remark in \cite{carlip} that the expression ($l=1,2$)
\begin{equation}
\int D[N^l]~e^{-i\int N^lH_l d^2z}, 
\end{equation}
if we want to respect invariance under diffeomorphisms has to be
understood as
$
\displaystyle{\delta(\frac{H_i}{\sqrt h})}
$
and similarly for $N$ and $H$.

Expression (\ref{reduced}) for the functional integral tells us
little about the ordering 
problem. 
In the case of two particles (${\mathcal N}=2$), the choice performed
in \cite{CMS1} was dictated by naturalness and 
aesthetic reasons reaching the logarithm of the Laplace-Beltrami
operator on a cone. As discussed in \cite{CMS1} this is very similar
to the quantum treatment of a test particle on a cone given in
\cite{deserjackiwcmp}. But there is no a priori reason for that
choice. A 
standard choice for the functional integral is the mid point rule
which is equivalent to the Weyl ordering at the operator
level \cite{lee}. In our case
\begin{equation}
H=\ln \left[(q\bar q)^{\mu_0} P\bar P\right] = \ln
\left[(q_1^2+q_2^2)^{\mu_0}(P_1^2+P_2^2)\right]  
\end{equation}
and the Weyl ordering gives rise simply to the operator 
\begin{equation}
\mu_0\ln(\hat{q}_1^2+\hat{q}_2^2)+\ln(\hat{P}_1^2+\hat{P}_2^2).
\end{equation}
This is the choice examined in
\cite{ciafaloni} in the context of high energy behavior of Yang-Mills
field theory. 
It appears that the logarithm of the Laplace-Beltrami
operator can be obtained only through a rather complicated ordering
process and the same can be said for the functional translation of the
Maass laplacian adopted in \cite{hosoya,carlip2}. For more than two
particles the hamiltonian even though perfectly defined, becomes very
complicated \cite{CMS1} and here up to now 
no guiding principle has emerged for addressing the ordering
problem.


\begin{thebibliography}{0}

\bibitem{DJH} A. Staruszkiewicz, Acta Phys. Polonica 24 (1963) 734; 
S. Deser, R. Jackiw and G. 't Hooft, Ann. Phys. (NY) 152
(1984) 220; S. Deser and R. Jackiw, Ann. Phys. 153 (1984) 40.
G. 't Hooft, Class. Quantum Grav. 9 (1992) 1335; {\it ibid.} 10 (1993)
1023, 10 (1993) 1653, 13 (1996) 1023.

\bibitem{carlip} S. Carlip, Class. Quant. Grav. 12 (1995) 2201.

\bibitem{cantinimenotti}L. Cantini, P. Menotti,
 Class. Quant. Grav. 20 (2003) 845 and references therein.

\bibitem{BCVW} A. Bellini, M. Ciafaloni, P. Valtancoli, Physics
Lett. B 357 (1995) 532; Nucl. Phys. B 454 (1995) 449; Nucl. Phys. B
462 (1996) 453; M. Welling, Class. Quantum Grav. 13
(1996) 653; Nucl. Phys. B 515 (1998) 436.

\bibitem{MS} P. Menotti, D. Seminara, Ann. Phys. 279 (2000) 282;
Nucl. Phys. (Proc. Suppl.)  88 (2000) 132.

\bibitem{CMS1} L. Cantini, P. Menotti, D. Seminara,
Class. Quant. Grav. 18 (2001) 2253; Nucl. Phys. B 638 (2002) 351.   

\bibitem{henneaux} M. Henneaux, C. Teitelboim, ``Quantization of gauge
systems'', Princeton University Press, Princeton (1992).


\bibitem{dhoker} E. D'Hoker, S. B. Giddings, Nucl. Phys. B 291 (1987)
90; E. D'Hoker, D. H. Phong, Rev. Mod. Phys. 60 (1988) 917.

\bibitem{alvarez}  O. Alvarez, Nucl. Phys. B 216 (1983) 125;
J. Polchinski, Comm. Math. Phys. 104 (1986) 37; Z. Bern, E. Mottola,
S. K. Blau, Phys. Rev. D 43 (1991) 1212.

\bibitem{hawkinghunter} S.W. Hawking and C. J. Hunter, Class. Quantum
Grav. 13 (1996) 2735.

\bibitem{deserjackiwcmp} S. Deser, R. Jackiw, Comm. Math. Phys. 118
(1988) 495.

\bibitem{lee} See e.g. T. D. Lee, ``Particle physics and introduction
to field theory'', Harwood  Academic Publishers, Switzerland (1988).

\bibitem{ciafaloni} M. Ciafaloni, S. Munier, Phys. Lett. B544 (2002) 307. 

\bibitem{hosoya}  A. Hosoya, K. Nakao, Prog. Theor. Phys. 84 (1990) 739.

\bibitem{carlip2}  S. Carlip, Phys. Rev. D 42 (1990) 2647;
Phys. Rev. D 45 (1992) 3584.

\end{thebibliography}
\end{document}